\documentclass{amsart}[12pt]

\usepackage[figuresright]{rotating}

\usepackage{graphicx}
\usepackage{bm}
\usepackage{amsmath}
\DeclareMathOperator*{\argmax}{arg\,max}

\newtheorem{theorem}{Theorem}%[section]

\newcommand{\defeq}{\buildrel \text{d{}ef} \over =}

\usepackage{algorithmicx}
\usepackage{algorithm}
\algnewcommand\algorithmicinput{\textbf{Input:}}
\algnewcommand\Input{\item[\algorithmicinput]}
\algnewcommand\algorithmicinitilize{\textbf{Initilize:}}
\algnewcommand\Initilize{\item[\algorithmicinitilize]}
\algnewcommand\algorithmicoutput{\textbf{Output:}}
\algnewcommand\Output{\item[\algorithmicoutput]}
\algnewcommand\algorithmicalg{\textbf{Algorithm:}}
\algnewcommand\Algorithm{\item[\algorithmicalg]}
\algnewcommand\algorithmicnote{\textbf{Note:}}
\algnewcommand\Note{\item[\algorithmicnote]}

\begin{document}

\title[Optimal Design for Change-Point Models]{Optimal Study Design for Reducing Variances of Coefficient Estimators in Change-Point Models}
%\pubyear{2020}
%\artmonth{April}
\author{ Li Xing $^{1,4}$,  Xuekui Zhang $^{2, *}$, Ardo van den Hout  $^{3}$,  \\Scott Hofer $^{4}$, Graciela Muniz Terrera $^{5}$}
\address{$^{\text{\sf 1}}$ Department of Mathematics and Statistics, University of Saskatchewan, Saskatoon, SK, S7N 5E6, Canada\\
$^{\text{\sf 2}}$ Department of Mathematics and Statistics, University of Victoria, Victoria, BC, V8N 1Y2, Canada\\
$^{\text{\sf 3}}$ Department of Statistical Science, University College London, London, WC1E 6BT, U. K. \\
$^{\text{\sf 4}}$ Institute on Aging \& Lifelong Health, University of Victoria, Victoria, BC, V8N 1Y2, Canada\\
$^{\text{\sf 5}}$ Centre for Dementia Prevention, The University of Edinburgh, Edinburgh, EH16 4UX, U.K.\\
$^{\text{\sf *}}$ Corresponding author: Xuekui@UVic.ca}
\email{Xuekui@UVic.ca}

\date{\today}

\begin{abstract}

In longitudinal studies, we observe measurements of the same variables at different time points to track the changes in their pattern over time. In such studies, scheduling of the data collection waves (i.e. time of participants’ visits) is often pre-determined to accommodate ease of project management and compliance. Hence, it is common to schedule those visits at equally spaced time intervals. However, recent publications based on simulated experiments indicate that the power of studies and the precision of model parameter estimators is related to the participants’ visiting schemes. 

In this paper, we consider the longitudinal studies that investigate the changing pattern of a disease outcome, (e.g. the accelerated cognitive decline of senior adults). Such studies are often analyzed by the broken-stick model, consisting of two segments of linear models connected at an unknown change-point. We formulate this design problem into a high-dimensional optimization problem and derive its analytical solution. Based on this solution, we propose an optimal design of the visiting scheme that maximizes the power (i.e. reduce the variance of estimators) to identify the onset of accelerated decline. Using both simulation studies and evidence from real data, we demonstrate our optimal design outperforms the standard equally-spaced design.

Applying our novel design to plan the longitudinal studies, researchers can improve the power of detecting pattern change without collecting extra data.     

\end{abstract}
\maketitle
\pagestyle{plain}

\section{Background}

There is sufficient evidence that demonstrates that biological, neuroanatomical, and behavioral changes occur prior to clinically identifiable symptoms and diagnosis of dementia and also prior to death \cite{Karr:2018cl}. Improving the accuracy and early identification of such changes are key current research targets in research on aging and dementia \cite{Jack2010}.  Longitudinal studies provide opportunities to investigate the onset of accelerated decline, particularly when there is sufficient information prior to change-points. In most of these longitudinal studies, however, the scheduling of the data collection waves (i.e. time of participants' visits) is often pre-determined to accommodate ease of project management \cite{Youngblut:wv} and compliance \cite{Cotter:2005by, doi:10.1177/001872676501800407}, most typically with data collected at equal and relatively widely-spaced intervals \cite{Neal:2011ff}. However, alternative designs have also been proposed, including a time-lag design to control for retest effects \cite{McArdle:1997} and cohort-sequential designs to compare individuals of the same age but born in different historical periods.

In recent publications, \cite{Rast:2014ej} and \cite{Timmons:uk} conducted simulation experiments to show that some designs improve study power and precision of model parameter estimation in a linear mixed-effects model framework. Following this work, we would like to investigate an optimal design to maximize the power to identify the accelerated decline. More specifically, the optimal study design in this content is a time scheduling scheme of patient visits, which maximizes the study power (or equivalently, minimize the variance of the model parameter estimators) given a fixed number of visits per participant. Different from other literature, our focus is finding the optimal design within the change-point model framework, as the modeling approach is commonly used to estimate the onset of accelerated change as discussed next \cite{Terrera2011}. 

Change-point models \cite{Yu:2012hj, Gadda:2006gb, Hall:2000cu} are used to model stochastic process changes over time. \cite{vandenHout:2011hy} summarized three commonly used change-point models. Among them, the broken-stick model, i.e. two linear segments joining at a change-point, is simple yet useful with a clear interpretation at all the model parameters, and can provide decent estimation for model parameters using fewer follow-up visits for each participant. Therefore, we would like to use the broken-stick model for data with an underlying change-point pattern. Particularly, the question that concerns us is when using the broken-stick model,  what is the optimal way to schedule visits in order to achieve better power. This is an important question, but not yet discussed in literature, which motivates our research team, statisticians and psychologists, to work on it. 

In Section 2, we formulate this optimal study design into a high-dimension-optimization problem, whose computation is too intensive to get a direct solution. However, by imposing reasonable constraints, we obtain a theoretical closed-form solution and provide an algorithm to transform the theoretical conclusion into practical procedures on data collection. Afterwards, we illustrate this optimal design through a few examples. In Section 3, we demonstrate that our proposed optimal design outperforms the traditional equally-spaced design via simulation studies. In Section 4, we show that the optimal design improves the power of study (i.e. precision of parameter estimators) in real data analysis. We provide discussion and conclusion in Section 5.

% This task is a high dimensional optimization problem since hypothetically the collection occasions of a subject as a time sequence can happen any time during the study. Through simulation studies and a real data application, we demonstrate that our optimal design works better than the commonly used equally-spaced design in most cases.              

\section{Methods and Examples}
\subsection{Assumptions, notation, and formulation of the optimization problem}

Without loss of generality, we simplify the discussion by focusing on one single patient and ignore the patient level index for the time being. Assume that there are $n$ visits scheduled for a patient. Let the visit time for the $j$th visit be $x_{j}$ and the change-point be $\gamma$, where $j = 1, 2,\cdots, n$ and $0=x_{1} \leq \cdots \leq x_{\tau} \leq \gamma \leq x_{\tau + 1}\leq \cdots \leq x_{n} =1 $. Note that the times of visits are scaled into the range of $[0,1]$ to simplify the discussion.  Denote $y_{j}$ as the cognitive score measured at time $x_{j}$. The broken-stick model is written as follows  
\begin{eqnarray}\label{model1}
y_{j} &=& \beta_{0} + \beta_{1} x_{j}  + \beta_{2}(x_{j}- \gamma)_{+}(x_{j}- \gamma) + \epsilon_{j},
\end{eqnarray}  
where \
\begin{equation}
\nonumber (x_{j} - \gamma)_{+} = \left\{ \begin{array}{cc}
0  & \text{if  }  x_{j} < \gamma \\
1  & \text{if  }  x_{j} \geq \gamma, 
\end{array}\right.
\end{equation}
and $\epsilon_{j} \buildrel \text{i{}.i.d} \over \sim \mbox{N}(0, \sigma^{2})$ is the random noise. Given a prior distribution of the change-point $\gamma$, $p(\gamma)$, which is usually obtained by historical data, we have the marginal likelihood function 
$$L(\bm \beta, \sigma^{2} | \bm Y, \bm X) = \int_{\gamma} L_\gamma(\bm \beta, \sigma^{2} | \bm Y, \bm X_{\gamma}) p(\gamma )d\gamma,$$ 
where the conditional design matrix 
\begin{equation}
\nonumber \bm{X}_{\gamma} \defeq \left(
\begin{array}{ccc}
1 & x_{1} & 0\\
\cdots & \cdots & \cdots\\
1 & x_{\tau} & 0\\
1 & x_{\tau + 1} & x_{\tau + 1} - \gamma \\
\cdots & \cdots & \cdots\\
1 & x_{n} & x_{n} - \gamma \\
\end{array}
\right).
\end{equation}
And for a given $\gamma$, we have the conditional likelihood function 
\begin{equation}
L_\gamma (\bm \beta, \sigma^{2} | \bm X_{\gamma}, \bm Y) = \mbox{C} \times \mbox{exp}\left(-\frac{1}{2\sigma^{2}}\left(\bm{Y} - \bm{X}_{\gamma}\bm{\beta}\right)^{T}\left(\bm{Y} - \bm{X}_{\gamma}\bm{\beta}\right)\right),
\end{equation}
where $\mbox{C}$ is a constant. Therefore, given a $\gamma$, the standard point estimates for $\bm \beta$ and $\sigma^{2}$ are 
\begin{align}
%\label{profilelik} l_\gamma(\hat{\bm\beta}_{\gamma}, \hat{\sigma}_{\gamma}^{2}) &= N(\bm{Y} ; \bm{X}_{\gamma}\hat{\bm{\beta}}_{\gamma}, \sigma^2 I_n)\\
\label{betagamma} \hat{\bm{\beta}}_{\gamma} &= (\bm{X}_{\gamma}^{T}\bm{X}_{\gamma})^{-1}\bm{X}_{\gamma}^{T}\bm{Y},\\
\label{sigmagamma} \hat{\sigma}_{\gamma}^{2} &=(\bm{Y} - \bm{X}_{\gamma}\hat{\bm{\beta}}_{\gamma})^{T}(\bm{Y} - \bm{X}_{\gamma}\hat{\bm{\beta}}_{\gamma})/(n-3).
\end{align}
As for the prior of $\gamma$, we could borrow information from historical data for estimating the distribution of the change-point location (i.e. the onset time of accelerated cognitive decline). Also, the prior information can be patient specific, which might be a function of age, gender, and other personal characteristics. Therefore, it can introduce the personal specific design on the visiting scheme according to the personal specific prior. In addition, in the practice of clinical studies, time is recorded as a discrete measurement with a minimum unit such as day and hour or just day. Therefore, in this work we assume time is a finite discrete measurement. That is, the values of the visit times, $x_j$'s, and the change-point, $\gamma$, are in a finite set of discrete numbers. Particularly, $\{\gamma_{1}, \ldots, \gamma_{K}\}$ listed in the increasing order are all possible locations of $\gamma$. We denote the probability mass function (PMF) of the distribution of prior knowledge about change-point location as \begin{align}
\label{prior} \mbox{Pr}(\gamma = \gamma_{k} ) = p_k,  \mbox{ and }\sum_{k = 1}^{K} p_{k} = 1 . %\mbox{ for } k = 1, 2, \cdots, K,
\end{align}
The prior PMF is usually estimated from the literature or provided by field scientists. Even if the prior knowledge is represented by a continuous distribution, we can always discretize it by binning. 
 
Our optimal design targets an optimal visiting scheme, which minimizes the variances of the estimated coefficients $\hat{\bm \beta}$ under a prior knowledge of the change-point $\gamma$. By minimizing the variances, we expect increasing accuracy and power. Because the variances of multiple parameters cannot be minimized simultaneously, in practice we minimize a pre-defined linear combination of variances of all regression parameters $ \hat{\bm \beta}$. Therefore, we formulate the design problem into an optimization question. That is, given the prior distribution on $\gamma$, 
\begin{align}\label{eq:optim0}
\argmax_{x_1,\ldots,x_n } \left( a_0 \mbox{Var}(\hat{\beta}_0) + a_1 \mbox{Var}(\hat{\beta}_1) +a_2 \mbox{Var}(\hat{\beta}_2) \right),
\end{align} 
where $a_0$, $a_1$, and $a_2$ are non-negative weights. There are some remarks on the weights. First, in practice we always set $a_0=0$ since most time researchers are interested in the rate of cognitive decline before the change-point, $\beta_1$, and the accelerated decline after the change-point, $\beta_2$. Second, the hypothesis testing with $\mbox{H}_{0}: \beta_2=0$ versus $\mbox{H}_{1}: \beta_2 \neq 0$ is equivalent to test if there is no decline increment after $\gamma$ or not, or say, if the change-point exists or not. Therefore, we specify the weights as $(a_0, a_1, a_2)=(0,0,1)$ to increase the signal for testing if the change-point exist or not.  Third, to increase the accuracy on estimating the cognitive decline rates both before and after the change-point, we specify the equal weights $(a_0, a_1, a_2)=(0,1,1)$, or the unequal weights $(a_0, a_1, a_2)=(0,1,2)$ to represent different focus on two decline rates.

\subsection{Derivation of the approximate solution for the optimization problem}

First of all, there is no closed-form solution for the optimization problem~(\ref{eq:optim0}). Second, for a numerical calculation, although the finite-discrete-time assumption refines the solution of this optimization problem in finite spaces, for a large $n$, it is impossible to conduct the brute-force searching due to the high dimensionality of the solution space. Our strategy is to relax a constraint of this optimal problem, which helps convert the problem into a problem with a closed-form solution. And the closed-form solution for the converted problem is a reasonable approximate solution for the original problem. The relaxation is to allow assigning proportions of the total number visits to given locations, where these proportions correspond to non-integer numbers of visits. For example, in the relaxed condition, we allow assigning $1.5$ visits on the $35$-th day of a clinical study. Since $1.5$ visits is impossible in practice, later we propose an algorithm on how to schedule those non-integer numbers of visits in practice.   

In the rest of this subsection, our discussions focus on how to convert the optimization problem after relaxing this constraint, and how to derive the closed-form solution for the converted problem. And we start this discussion under a no change-point scenario, extend it to the situation under a known change-point, and then generalize it to the common situation of unknown change-point locations.
 
\subsubsection{No change-point} When there is no change-point, the broken-stick model degenerates into a simple linear regression model. In this case, $\mbox{Var}\left(\hat{\bm \beta}\right)  = \sigma^{2}\left(\bm{X}^{T}\bm{X}\right)^{-1},$ where $\left(\bm{X}^{T}\bm{X}\right)^{-1}$ is a $2 \times 2$ matrix with diagonal elements proportional to the variances of the intercept estimator and the slope estimator correspondingly. Specifically, $\mbox{Var}(\hat{\beta}_{1}) \propto 1/\sum_{j=1}^{n}\left(x_{j} - \bar{x}\right)^{2}$, which is minimized when the time of visits are evenly scheduled at the two boundaries ($0$ and $1$).

\subsubsection{A known change-point} Denote the known change-point as  $\gamma_{1}$. The broken-stick model becomes two segmented lines connected at $\gamma_{1}$. And the boundaries of the two segments are $(0, \gamma_{1})$ and $(\gamma_{1}, 1)$, respectively. According to the result of the no-change-point situation, we claim that all visits must be scheduled at their boundaries, $(0, \gamma_{1}, 1)$, to obtain the minimum variances of the two slope estimators from the two segments. The proof is straight forward. Assume that a proportion of the visits, $q_s$, is assigned to some non-boundary point, $s$, in the second segment. We would like to split this proportion into half, and assign these two halves to the two boundary points $\gamma_{1}$ and $1$ separately. Such re-assignment improves the slope estimate in the second segment (according to the result of the no-change-point situation), and also improves the slope estimate in the first segment since the proportion of visits added to $\gamma_{1}$ help to reduce variance in the slope estimation for the first segment. Therefore, the final solution can be denoted by $q_{0}, q_{1}, q_2$, which correspond to the proportions of visits (i.e. $x_j$'s) assigned to the three locations $0, \gamma_{1}, 1$, respectively, with $q_{0} + q_{1} + q_{2} = 1$. And since every study design under a known change-point condition corresponds to a set of values $(q_0, q_1, q_2)$, we reduce the dimension of the optimization problem from $n$ to $3$ for the known $\gamma$ condition. The three variance terms in problem~(\ref{eq:optim0}) can be written as functions of $(q_0, q_1, q_2)$ as follows %To solve the optimization problem, we write the variances of the coefficient estimates as functions of $
\begin{align}
\mbox{Var}(\hat{\beta}_{0} |\gamma_{1}) &=\frac{\sigma^{2}}{nq_{0}}, \label{eq:beta0}\\ 
\mbox{Var}(\hat{\beta}_{1}|\gamma_{1}) &= \frac{\sigma^{2}}{\gamma^{2}_{1}}\left(\frac{1}{nq_{0}} + \frac{1}{nq_{1}}\right), \label{eq:beta1} \\ 
\mbox{Var}(\hat{\beta}_{2}|\gamma_{1}) &= \frac{\sigma^{2}}{\gamma_{1}^{2}(1-\gamma_{1})^{2}} \left(\frac{(1-\gamma_{1})^{2}}{nq_{0}} + \frac{1}{nq_{1}} + \frac{\gamma_{1}^{2}}{nq_{2}} \right). \label{eq:beta2}
\end{align}
The detailed derivation of Equations~(\ref{eq:beta0}) to (\ref{eq:beta2}) is provided in Lemma 1 of the web appendix. In equation~(\ref{eq:beta0}), $\mbox{Var}\left(\hat{\beta}_{0}\right) \propto \frac{1}{q_{0}}$, the inverse of the proportion at $0$, which tells the more visits assigned at time $0$ the more reduction in variance of the intercept estimate. In equation~(\ref{eq:beta1}), $\mbox{Var}\left(\hat{\beta}_{1}\right) \propto \frac{1}{q_{0}} + \frac{1}{q_{1}} $, which tells we reduce the variance of the slope estimate in the first segment by assigning more visits at $0$ and $\gamma_{1}$. Also, $\mbox{Var}\left(\hat{\beta}_{1}\right) \propto \frac{1}{\gamma^{2}_{1}}$, which tells that it is harder to precisely estimate $\beta_{1}$ if the change-point $\gamma_{1}$ happens earlier. Equation~(\ref{eq:beta2}) shows that the variance of $\hat{\beta}_{2}$ is regulated by all model parameters ($\gamma_{1}, q_{0}, q_{1}, q_{2}$). In summary, parameters ($q_{0}, q_{1}, q_{2}$) contribute to the variances in form of linear combinations of their inverses. 

\subsubsection{Unknown change-point}
When the change-point is unknown, the PMF of $\gamma$~(\ref{prior}) provides the prior knowledge of the change-point locations. Based on this PMF, we convert the unknown change-point problem into a weighted combination of $K$ known-change-points problem, and calculate the variance of the estimated coefficients ($\hat{\beta}$'s) based on ensembling information from these known change-point problems. The detailed mathematical derivation of the variances is provided in Lemma 2 of the web appendix.
%    Consider a clinical study last $K+1$ days, and the change-point can be at any day except the first and last day of study. We normalize the increasingly ordered time of this study to interval $[0,1]$ as $(\gamma_{1}=0 , \gamma_{1}, t_{2},  \ldots , \gamma_{k} ,\gamma_{K+1}=1)$. Let the $k$-th possible change-point location to be the $k$-th non-boundary time point, i.e.  $\gamma_{k}=\gamma_k$. We denote the probability mass function (PMF) for prior distribution of change-point as %For this condition, we assume a probability mass function (PMF) for the change-point location, which can be specified based on the prior knowledge from the literature. This PMF is denoted as 
%    \begin{align}
%    \label{prior} \mbox{Pr}(\gamma = \gamma_{k} ) = p_k,  \mbox{ and }\sum_{k = 1}^{K} p_{k} = 1  %\mbox{ for } k = 1, 2, \cdots, K,
%    \end{align}
%    Note that even if the prior knowledge is represented by a continuous distribution, we can always discretize it by binning. 
Denote $q_0, q_1, \ldots, q_K, q_{K+1}$ as the proportions of visits assigned to time, $0, \gamma_{1}, \ldots, \gamma_{K}, 1$, separately. The results of Equations~(\ref{eq:beta0}) to (\ref{eq:beta2}) are now extended into the linear combinations of $1/q_k$'s as follows
\begin{align}
\label {eq:lem3.1} &\mbox{Var}(\hat{\beta}_{0}) =c_0 + \frac{\sigma^{2}}{nq_{0}},\\ 
\label {eq:lem3.2} &\mbox{Var}(\hat{\beta}_{1}) = c_1 + \sigma^2 \left(\frac{  \sum_{k = 1}^{K} A_{k}}{nq_{0}} + \sum_{k = 1}^{K} \frac{A_k}{nq_k} \right), \\ 
\label {eq:lem3.3} &\mbox{Var}\left(\hat{\beta}_{2}\right) =  c_2 +   \sigma^2 \left( \frac{ \sum_{k = 1}^{K} A_{k}}{nq_{0}} 
												+ \sum_{k = 1}^{K} \frac{A_kB_{k}}{p_{k}nq_{k}} 
												+   \frac{ \sum_{k = 1}^{K} B_{k}}{nq_{K+1}} \right),\\
\label{eq:lem3.4}&A_{k} = \frac{p_{k}}{\gamma_{k}^{2}} ,\;\;\;  B_{k} = \frac{p_{k}}{(1-\gamma_{k})^{2}},
\end{align}
where $(c_0, c_1, c_2) = \mbox{diag}\left( \mbox{Var}_{\gamma}\left(\bm{\beta}_{\gamma}\right)\right)$, a vector of the unknown truth, is a constant with respect to the design parameters $q_k$'s. Plugging the Equations~(\ref{eq:lem3.1}) into (\ref{eq:lem3.4}) to the original optimization problem~(\ref{eq:optim0}), we have a new optimization problem with respect to $(q_{0}, q_1, \dots, q_K, q_{K+1})$, which has a closed-form solution provided in Theorem~\ref{theorem2}. (The detailed proof is provided in the web appendix.) 
\begin{theorem}\label{theorem2} 
Given the constraint $\sum_{k=0}^{K+1}q_k = 1$, the values of $q_k$'s that minimize $a_0 \mbox{Var}(\hat{\beta}_0) + a_1 \mbox{Var}(\hat{\beta}_1) +a_2 \mbox{Var}(\hat{\beta}_2)$ are 
\begin{equation}
q_k =  \sqrt{d_k} / \sum_{k=0}^{K+1} \sqrt{d_k}, 
\end{equation}
where $d_0 = a_0 + (a_1+a_2) \sum_{k=1}^K A_{k} $, $d_{k} = a_1 A_k + a_2 A_k B_{k}/p_{k}$, $d_{K+1}  =a_2  \sum_{k=1}^K B_{k}$ and $A_{k}$ and $B_{k}$ as defined in (\ref{eq:lem3.4}).
\end{theorem}
The solutions of $q_k$'s in Theorem~\ref{theorem2} are used to design the optimal visit scheme, which is discussed in the next subsection.
%
%Based on lemma 1 in the appendix, the variance of estimated coefficients $\mbox{Var}\left(\hat{\bm \beta}\right)$ can be divided into two parts, $\mbox{Var}_{\gamma}\left(\bm{\beta}_{\gamma}\right)$ and $\sigma^{2}\mbox{E}_{\gamma}\left(\left(\bm{X}_{\gamma}^{T}\bm{X}_{\gamma}\right)^{-1}\right)$.  The first term is invariant to the design. Only the second term $\mbox{E}_{\gamma}\left(\left(\bm{X}_{\gamma}^{T}\bm{X}_{\gamma}\right)^{-1}\right)$ can be controlled by study designs. Therefore, to minimize the linear combination of variances of estimated coefficients $\hat{\bm{\beta}}$ based on designs, only $\mbox{E}_{\gamma}\left(\left(\bm{X}_{\gamma}^{T}\bm{X}_{\gamma}\right)^{-1}\right)$ can be manageable.

\subsection{ The time scheduling scheme of the optimal design}\label{sec:assignT}
The scheduling scheme of the optimal design is usually not directly obtained from the solutions of $q_k$'s in Theorem~\ref{theorem2}  since $nq_k$ might be non-integer for some $k$'s. Especially, when $K >> n$, which is common in practice, $nq_k$'s are almost never larger than $1$.  For example, assume that we need to design a clinical study with $10$ visits per patient during one year period. And we use day as the time unit. The prior of the change-point tells that probability of the change-point is uniform during the year. In this case, we require $nq_0+ nq_1+ \ldots + nq_{366} =10$, and we know the majority of $nq_k$'s must be non-integer between $0$ and $1$. This is caused by relaxing the constraint, which we discussed at the beginning of this section. That is, we allow non-integer visits to obtain the closed-form solution for high dimensional optimization problems. After obtaining these optimal solutions, we need to figure out a scheduling scheme to make the optimal design useful in practice. Solutions of Theorem~\ref{theorem2} suggest assigning $q_k$ proportion of visits at time $\gamma_{k}$, for $k = 0, 1, \ldots, K+1$, with $\sum_k q_k = 1$. We construct a generating distribution of the optimal visiting scheme with PMF as $P(T = \gamma_{k}) = q_k$. Then its cumulative distribution function (CDF) is $F_T(\gamma_{k}) =P(T \leq \gamma_{k})=  \sum_{s=0}^k q_s$, which is a step function with $q_k$ increment at $\gamma_{k}$ for all $k$. 

Since in practice we can only assign integer numbers as the numbers of visits, a practical optimal design is required. Therefore, we propose a scheme generation distribution, whose CDF is a step function with the enforced increments proportional to $1/n$ at all $\gamma_{k}$'s given the total $n$ visits, and we suggest to assign the $j$-th visit at the $j/n$-th quantile of this CDF, for $j= 1, \ldots, n$. Since for discrete distributions, most of the times there are no solutions for equations $F_T(x_j) = P(T\leq x_j) = j/n$, instead of assigning the exact $j/n$-th quantile, we assign $x_j$ to the right-hand-side nearest neighbor corresponding to the $j/n$-th quantile at a given time of visit, i.e. 
\begin{align} \label{f:sol}
x_j = \gamma_{k},  \mbox{where } k=min \{t: F_T(\gamma_t) \geq j/n\} .
 \end{align}
Please note that we choose the right-hand-side neighbor, but not the left-hand side because the left-hand-side CDF is determined by history but not the newly assigned visit after that point. We provide examples to illustrate the idea of this decision in the subsection~\ref{sec:example}. In summary, the practical optimal design is based on assigning visits at the equal quantiles of scheme generating distribution. And therefore, we call it the EQ design. In contrast, we call the traditional design the ES design since it assigns visits at equally-spaced time points. And Algorithm~\ref{algm} summarizes the procedures on scheduling the optimal visiting time scheme in the EQ design.\begin{algorithm}[!hbtp]
	\caption{ Scheduling Scheme of the EQ Design}
	\label{algm}
	\begin{algorithmic}%[width=0.5\textwidth]
		\Input \\
		\textit{The number of visits to be scheduled:} $n$;\\
		\textit{A set of all possible visit time:} $\{0, \gamma_{1}, \ldots , \gamma_{K}, 1\}$;\\
		\textit{The weights for the variances of the three parameters in the broken-stick models:} $a_0, a_1, a_2$;\\
		\textit{The prior distribution of the change-point locations:} $\mbox{Pr}(\gamma=\gamma_{k})=p_k$ for $k=1,\ldots, K$.
			
		\Algorithm
		\State[\textbf{Step 1: Calculate quantiles $q_k$'s as the result of Theorem~\ref{theorem2}}] 
		\begin{align*}
                &\mbox{(Step 1.1) }  A_{k} = \frac{p_{k}}{\gamma_{k}^{2}} , \;\;\; B_{k} = \frac{p_{k}}{(1-\gamma_{k})^{2}};\\
%		\end{align*}
%		%\State[\emph{Calculate quantities defined in Lemma~\ref{lemma4}}]  
%		\begin{align*}
                &\mbox{(Step 1.2) }  \Rightarrow \left \{ \begin{array}{l} d_0 = a_0 + (a_1+a_2) \sum_{k=1}^K A_{k} ;\\
                                             d_{k} = a_1 A_k + a_2 A_k B_{k}/p_{k} \quad ( k=1, \ldots, K ); \\
                                             d_{K+1}  =a_2  \sum_{k=1}^K B_{k};
                       \end{array} \right. \\                    
                &\mbox{(Step 1.3) }   \Rightarrow q_k =  \sqrt{d_k} / \sum_{k=0}^{K+1} \sqrt{d_k}.\\
% 			d_0 &= a_0 + (a_1+a_2) \sum_{k=1}^K A_{k}, \;\;\;
%			d_{\gamma_k} = a_1 A_k + a_2 A_k B_{k}/p_{k}; \;\;\;
%			d_1 =a_2  \sum_{k=1}^K B_{k},\\
%			d^* &= \sqrt{d_0} + \sqrt{d_1} + \sum_{k=1}^K \sqrt{d_{\gamma_k}}\\
%			&q_0 &= \frac{\sqrt{d_0} }{d^*}, \;\;\; 
%			&q_{\gamma_k} = \frac{\sqrt{d_{\gamma_k}}}{d^*}, \;\;\;
%			q_{1} = \frac{\sqrt{d_1} }{d^*} 			
		\end{align*}					
		\State[\textbf{Step 2 : Calculate $x_j$'s for the optimal visiting scheme}]  \\
%		Construct a PMF, using possible visiting time $(0, \gamma_1, \ldots , \gamma_K,1)$ and their corresponding probabilities $(n_0/n, n_{\gamma_1}/n, \ldots , n_{\gamma_K}/n,n_1/n)$.
	         $x_j = \gamma_{k},  \mbox{where } k= min \{t: \sum_{s=0}^t q_s \geq j/n\}.$
		\Output \\
		The optimal visiting scheme $(x_1, \ldots, x_n)$.
		
%		\Note\\
%		For binary outcomes, replace $\hat{Y}$  by $\hat{p}=\mbox{P}(\hat{Y}=1)$, and replace $\tilde{Y}$ by $\tilde{p}=\mbox{P}(\tilde{Y}=1)$.
	\end{algorithmic}
\end{algorithm}

\subsection{Examples of the Optimal Design} \label{sec:example}

\subsubsection{An example with a known change-point location}
Under some special circumstance, the prior tells that the change-point location is at a given location, $\gamma_{1}$. That is, $\mbox{Pr}(\gamma = \gamma_{1}) = 1.$  
In this case, for the optimal study design, we assign visits only at the two boundary points and the change-point location, $0, \gamma_1, $ and $1$. After selecting $a_{0} = a_{1} = 0$ and $a_{2} = 1$ and then applying Theorem~\ref{theorem2}, we obtain the proportions of the assign visits as follows  
\begin{equation*}
q_0 = (1-\gamma_{1})/2, \quad q_{\gamma_1} = 1/2, \quad \mbox{and} \ q_1 = \gamma_{1}/2.
\end{equation*}
%This result is a straightforward application of Theorem~\ref{theorem2} by setting $a_{0} = a_{1} = 0, a_{2} = 1$.  In this case, . 
This result shows that (1) half of the visits need to be assigned to the known change-point, (2) more visits are suggested to be assigned to the boundary closer to the change-point than the one further away, and (3) the number of visits is proportional to the distance between the change-point to the opposite side boundary. The result makes sense since the longer distance introduces the larger variance in $x_j$'s, and also indicates easier to precisely estimate the regression slope. More specifically,  when $\gamma_1 = 0.2$ and $n = 10$,  we have $nq_0 = 3$, $nq_{_1} = 5$, and $nq_2 = 2$. As the numbers are all integers, assigning visit time $x_j$ is trivial. We simply set $x_1=x_2=x_3=0$, $x_4=x_5=x_6=x_7=x_8=0.2$, and $x_9=x_{10}=1$. 

Next we continue to use this example to illustrate the idea of using the right-hand-side nearest neighbor, but not using the left-hand-side nearest neighbor.  The CDF for the scheme generating distribution is a 3-step function defined as $\mbox{Pr}(t=0)=q_0=0.3$, $\mbox{Pr}(t \leq 0.2)=(q_0+q_{\gamma_1})=0.8$, and $\mbox{Pr}(t \leq 1)=1$. If we use left-hand-side nearest CDF values, the assignment is $x_1=x_2=x_3=x_4=x_5=x_6=x_7=x_8=0$, $x_9=0.2$, and $x_{10}=1$, which does not agree with the optimal scheme shown above, while the right-hand-side nearest neighbor provides the consistent result with the optimal design.   

\subsubsection{Examples with unknown change-point locations}
\begin{figure}[h!btp]
\begin{center}
%\begin{sidewaysfigure}
%    \centering
%    \vspace{-0.1in}
    \includegraphics[scale=0.6]{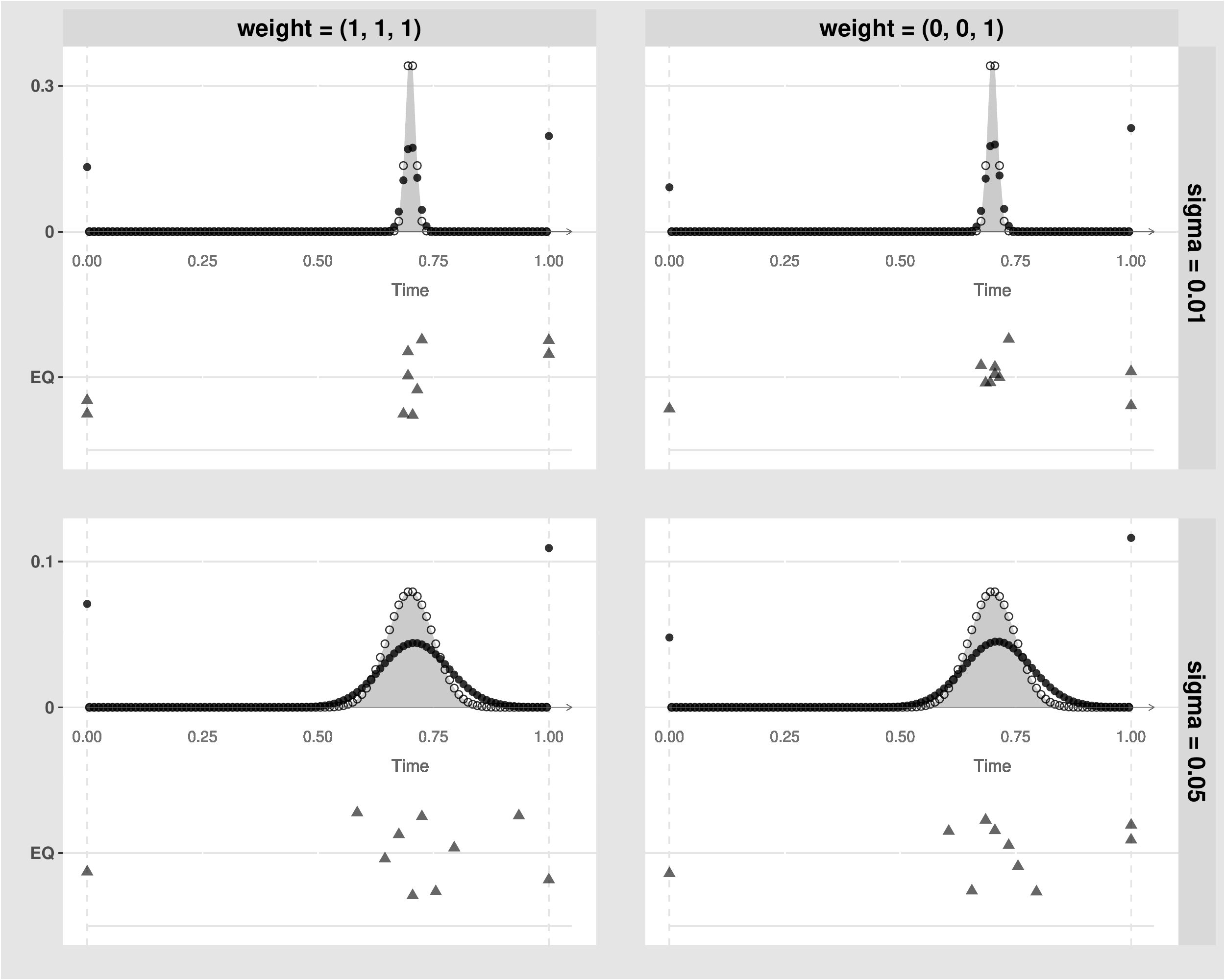}
    \caption{Four EQ Design Examples. In each panel, the shaded area represents the area under the density curve of prior distribution of the change-point. In the top row, prior distributions of the change-points are $\mbox{N}(0.7,0.01)$. In the bottom row, prior distributions of the change-points are $\mbox{N}(0.7,0.05)$. In the left column, weights considered in the EQ design are $(1,1,1)$. And, in the right column, weights are $(0, 0, 1)$. Black circles represent the discretized PMF from prior knowledge of the change-point locations, while solid dots represents the PMF of $x_j$'s in the optimal design. Triangles, jittered in $y$ direction, represent the exact values of $x_j$'s for the optimal design.}
    \label{fig:egDesign}
%\end{sidewaysfigure}
\end{center}
\end{figure}

The PMF of the optimal scheme generating distribution is always multimodal with two peaks on the boundaries. And other peaks of the PMF depend on the prior knowledge of the change-point. In this session, we use four cases to illustrate the shape of PMF and how the scheduling scheme works. The prior distribution of the change-point is \mbox{N}($0.7$, $\sigma$) with $\sigma =  0.01$ and $0.05$ indicating different confidence levels on the change-point location. The weights are either $(1, 1, 1)$ or $(0, 0, 1)$ emphasizing the different parameters of interests. The total number of visits is $n=10$. We first discretize the prior distribution of the change-point and then based on Algorithm~\ref{algm}, we have the scheme generating PMF and the practical optimal visiting scheme. Figure~\ref{fig:egDesign} provides a clear illustration on the four cases. There are peaks on two boundaries and the third peak on the mode of the change-point prior distribution. The non-boundary peak (solid dot) in the scheme generating PMF is always lower than the mode (circle) of the discretized prior distribution of the change-point location since we need to combine on the prior of change-point with information on the boundary and, therefore, the prior information is diluted or updated in the scheme generating PMF. Also, for the same reason, the scheme generating PMF is wider than the prior distribution of the change-point. The prior information of the change-point locations in the first row, $\mbox{N}(0.7, 0.01)$, is stronger than the one in the second row, $\mbox{N}(0.7, 0.05)$, which leads more points (triangles) assigned to the two boundaries to help improve estimations of the slopes in the first row. When we emphasize on minimizing $\mbox{Var}(\hat{\beta}_2)$ (in the right column), there are fewer points assigned in the left region from the left boundary to the mode of the change-point, compared with the cases when we optimize the average of all variances (in the left column).

%\begin{figure}[h!btp]
%\begin{center}
%\vspace{-0.1in}
%\includegraphics[scale=0.8,  angle =90]{mydesign.pdf}
%\caption{Examples of $4$ optimal designs. In the left column, the weights for the variances of all coefficient estimates are $(1,1,1)$. And, in the right column, the weights are $(0, 0, 1)$. In the top row,  the prior distribution of the change-point is $\mbox{N}(0.7,0.01)$. And, in the bottom row, the prior distribution of the change-point is $\mbox{N}(0.7,0.05)$. Black circles represent the discretized PMF from prior knowledge of change-point locations, while the red dots represents the PMF of $x_j$'s in the optimal design. The purple diamonds, which are jittered in the $y$ direction, represent the exact values of $x_j$'s for the optimal design.}  \label{fig:egDesign}
%\end{center}
%\end{figure}

\section{Simulation Studies}

We carry out simulation studies to compare the performance of the traditional ES design, which separates visits with the same-length time intervals,  with the proposed optimal EQ design, which assigns visits at the equal quantile of the PMF of the generating distribution. The generating distribution of the change-point is unknown in general and need to be estimated from the prior knowledge in the historical data. Therefore, we make comparisons under different prior assumptions and also evaluate the reliability of the EQ design under the mis-specified prior distributions. To make comprehensive comparisons, we identify important factors, vary the levels of these factors, and conduct a factorial experiment, which is described in the next subsection.  

\subsection{Simulation design and settings}
Let $y_{ij}$ be the cognitive scores for the $j$th subject at the $i$th visit with $j = 1, 2, \cdots, m$ and $i = 1, 2, \cdots, n$. The longitudinal data with $m$ subjects are generated using the model as follows  
\begin{equation}
y_{ij} = \beta_{0} + \beta_{1} x_{i}  + \beta_{2}(x_{i}- \gamma_{j})_{+}(x_{i}- \gamma_{j}) + a_{j} + \epsilon_{ij},
\end{equation}
where $\beta_{0} = 0, \beta_{1} = -2, \beta_{2} = -4$, $a_{j} \buildrel \text{i{}.i.d} \over \sim \mbox{N}(0, 0.01)$, and $\epsilon_{ij} \buildrel \text{i{}.i.d} \over \sim \mbox{N}(0, 0.01)$ with  $ m = 100$. We simulate the change-points, $\gamma_{j}$'s, from $\mbox{TN}(\mu, \sigma, 0, 1)$ indicating a normal distribution with center, $\mu$, and standard deviation, $\sigma$, truncated between the range of $[0, 1]$. And the three identified important factors include the total number of visits, $n$, and the mean, $\mu$, and the standard deviation $\sigma$ of the prior distribution. To cover all the possible change-point locations in simulations, we vary the center, $\mu$, as $0.25, 0.5$, and $0.75$, which represents an early change, a middle change, and a late change, respectively. We vary the standard deviation, $\sigma$, as $0.01,$ and $0.1$, which represents a confident belief and a less confident belief of the change-point locations across patients, respectively. Also we add mis-specified cases for the EQ design and denote them as EQ\_$80\%$ with prior $\mbox{TN}(\mu, 80\%\sigma, 0, 1)$, EQ\_$120\%$ with prior $\mbox{TN}(\mu, 120\%\sigma, 0, 1)$, EQ\_Left with prior $\mbox{TN}(\mu-0.02, \sigma, 0, 1)$ , and EQ\_Right with prior $\mbox{TN}(\mu+0.02, \sigma, 0, 1)$, which represents over-confidence or under-confidence of the prior distributions and the misspecified centers biased to the left or to the right, respectively. In this way, we get $30$ prior distributions by combination of all levels of the two factors for the EQ design. We also vary the total number of visits, $n$, as $10, 15, 20$ and $25$.  In summary, we have  $3$ ( $\mu$'s ) $ \times 2 $ ( $\sigma$'s ) $ \times 4 $ ( $n$'s ) $ \times 6$  (visit time schemes: one ES design, one EQ design and four mis-specified EQ designs) $ = 144$ settings.  For each setting, based on Algorithm~\ref{algm}, we get the times of visits, $x_{i}$'s, for the EQ designs, while those $x_{i}$'s for the ES design is just the boundaries and the separating points for equal intervals.      
%\begin{figure}[h!btp]
%\begin{center}
%\vspace{0.2in}
%\includegraphics[scale=0.8, angle =90]{size10V2.pdf}
%\caption{ Time of ten visits designed based on the ES design and the EQ design with different prior distributions. The blue diamonds represent the times of visits from the ES design. The orange diamonds represent the times of visits from the EQ design with $\mbox{TN}(\mu-0.02, \sigma, 0, 1)$. The green diamonds represent the times of visits from the EQ design with the correctly specified prior distribution, $\mbox{TN}(\mu, \sigma, 0, 1)$. The red diamonds represent the times of visits from the EQ design with $\mbox{TN}(\mu+0.02, \sigma, 0, 1)$.}  \label{fig:size10}
%\end{center}
%\end{figure}

\begin{figure}[h!btp]
\begin{center}
%\begin{sidewaysfigure}
%    \centering
%    \vspace{-0.1in}
    \includegraphics[scale=0.6]{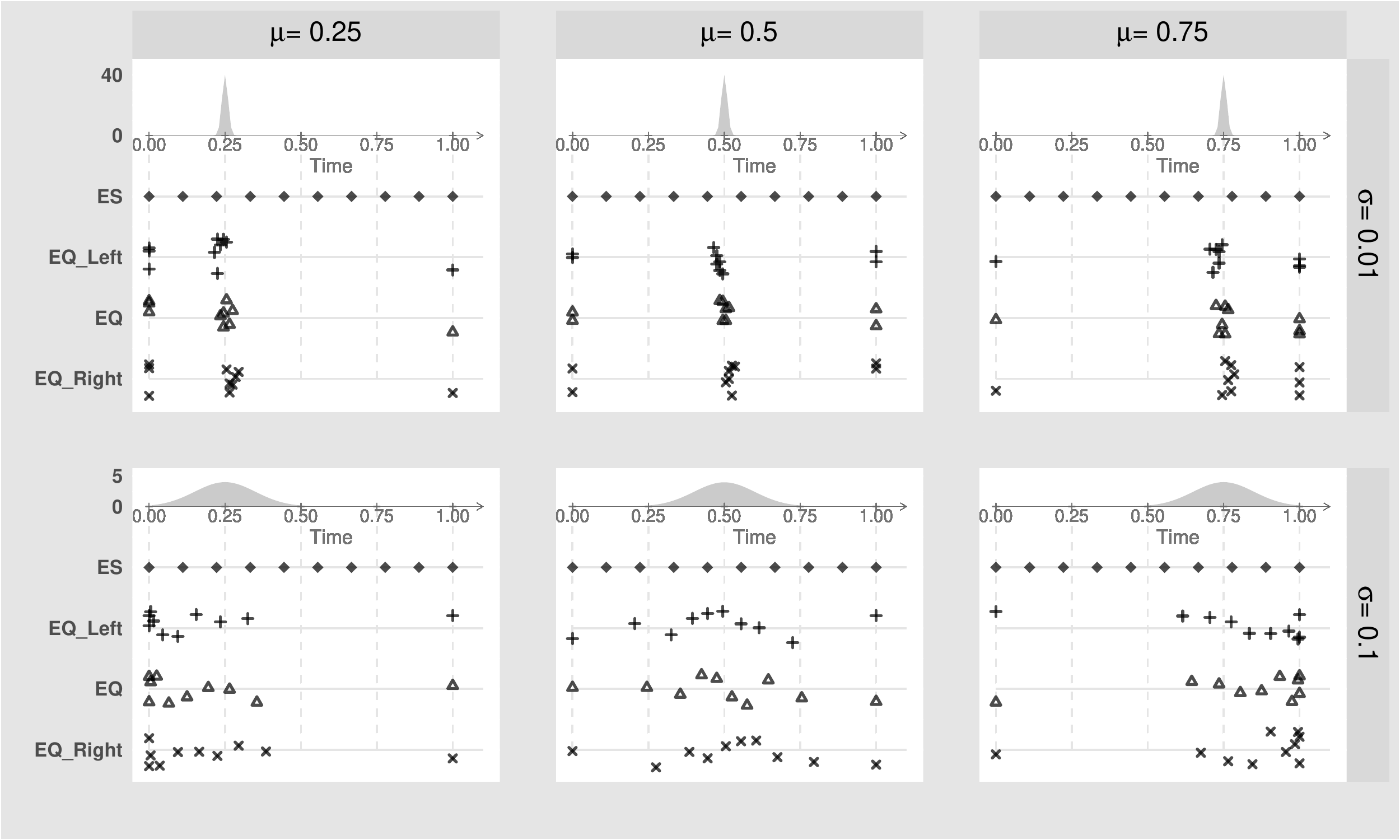}
    \caption{Visiting Schedules for Ten Visits Based on the ES or EQ Design under Different Prior Distribution Assumptions. In each panel, the shaded area indicates the densities of $ \mbox{TN}(\mu, \sigma, 0, 1) $, the prior distribution of the change-point.  The diamonds represent the visiting schedule based on the ES design. The triangles represents the visiting schedule based on the EQ design with the correctly specified prior distribution, $ \mbox{TN}(\mu, \sigma, 0, 1) $. The plus symbols represent the visiting schedule based on the EQ\_Left design, i.e. the EQ design with mis-specification to the left side ($ \mbox{TN}(\mu-0.02, \sigma, 0, 1) $). The cross symbols represent the visiting schedule based on the EQ\_Right design, i.e. the EQ design with mis-specification to the right side ($ \mbox{TN}(\mu+0.02, \sigma, 0, 1) $).}
 \label{fig:size10}
%\end{sidewaysfigure}
\end{center}
\end{figure}
Figure \ref{fig:size10} illustrates the differences in the scheduled visits under different settings at $n = 10$. The ES design uses the same equal space strategy under all settings, while the EQ design varies according to the given prior information.    

\subsection{Simulated data and analysis model} 

For each setting, we firstly obtain $x_{i}$'s for the ES design and the EQ designs separately. And then we simulate $100$ data sets based on those $x_{i}$'s as replications, which results a total of $14,400$ data sets. For each data set, we conduct the Bayesian inference of a change-point model using the Hamiltonian Monte Carlo algorithm \cite{Neal:2011ff}  built in Stan \cite{JSSv076i01}, which is also implemented in the R package, \textit{rstan}, to call Stan from R \cite{Rsoftware}. In the Bayesian inference, we specify a diffuse prior, $\mbox{N}(0, 100)$, for the regression coefficients,  $\beta_{0}, \beta_{1}, \beta_{2}$, a uniform prior, $\mbox{U}(0,1)$ for the change-points, and a diffuse prior $\mbox{U}(0.001, 100)$ for the standard deviation of the white noise. And the posterior standard deviations of the coefficients are used as measurement of the variability of the coefficient estimators. 

\subsection{Simulation results}
\begin{figure}[h!btp]
\begin{center}
\vspace{0.2in}
\includegraphics[scale=0.6]{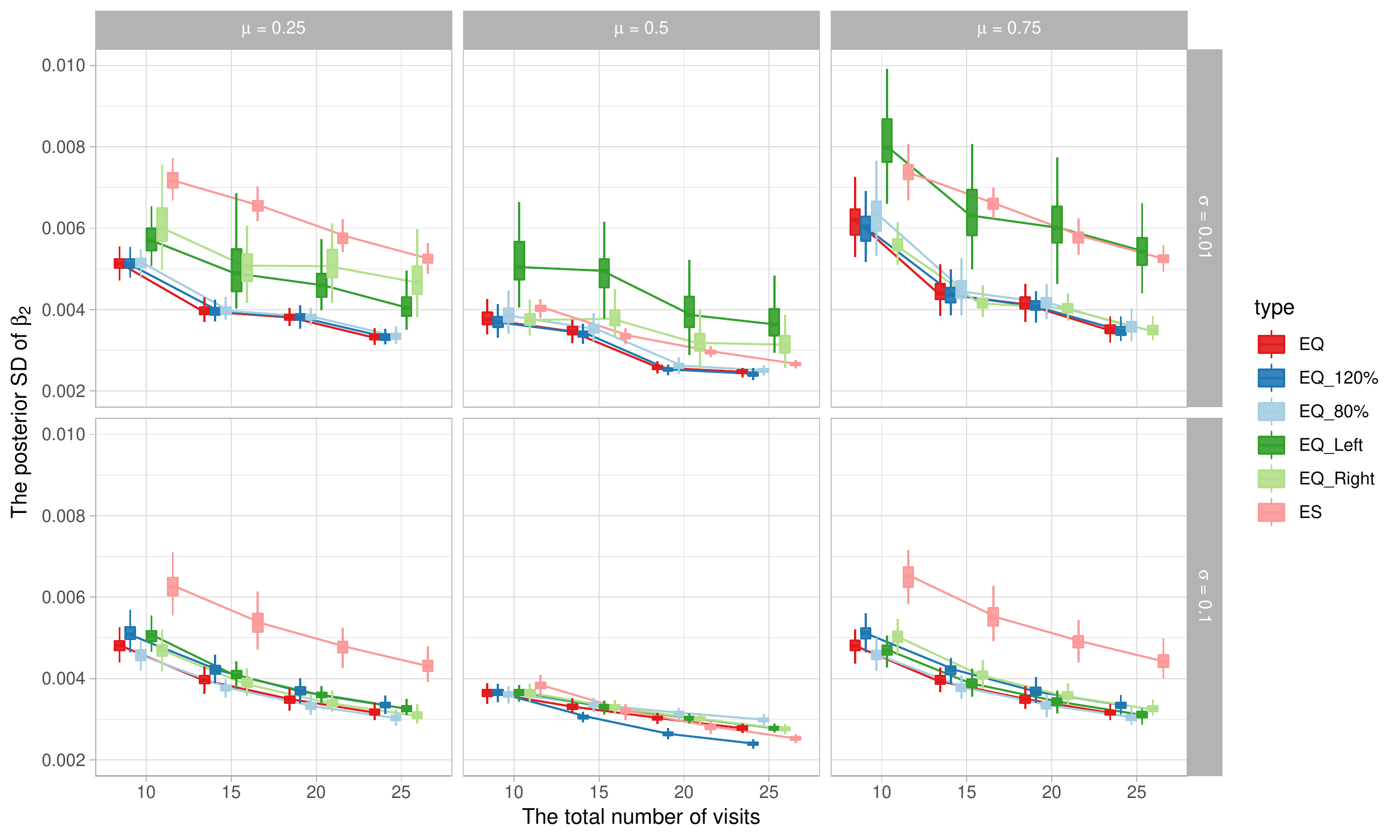}
\caption{Boxplots of the posterior standard deviations of $\beta_{2}$ under the total of 144 simulation settings.}  \label{fig:best}
\end{center}
\end{figure}
Figure \ref{fig:best} illustrates distributions of the posterior standard deviations of $\beta_{2}$ under different settings. The comparisons among the design schemes shows that, under the correctly specified prior distributions, the EQ design outperforms the ES design in most of the settings except under $\mu = 0.5$, $\sigma = 0.1$ and the total number of visits greater than $15$. The exceptions are due to three reasons. First, when the prior distribution is balanced ($\mu = 0.5$), the symmetry improves the performance of both the ES design and the EQ designs with the correct centers. And we see that the improvement on the ES design is more prominent.  Second, the larger the variance the similar performance between the ES design and the EQ design since when the underlying prior distribution becomes more and more flat, there is less information carried to the design, which eventually leads to the equally spaced times of visits under the non-informative prior. At last, since we generate the longitudinal data with random effects and white noise, it may also introduce some uncertainty.

For the mis-specified cases on the centers,  given the underlying prior distribution, $\mbox{TN}(\mu, \sigma, 0, 1)$ with $\sigma = 0.01$, the $0.02$ shift makes the mis-specified center outside the $95\%$ probability region of the change-point, which is a severe mis-specification. Therefore, the ES design outperforms these two kinds of mis-specification (the left and right shifts) in most cases when $\mu = 0.5$. However, when the true center is at $0.25$, these two kinds of mis-specification EQ designs still outperform the ES design in most cases. When the true center is at $0.75$, the mis-specification on the right shift outperforms the ES design, but the mis-specification to the left performs similarly compared with the ES design, but with a larger variation. This tells us that in practice if we know that the change-point usually happens in a late stage, it is safer to put more points close to $1$.            

For the mis-specified cases on the standard deviations, the EQ designs with mis-specification have the similar performance with the EQ design without mis-specification, which demonstrates the robustness of the EQ design on the misspecification of the variance. We also notice that at $\mu = 0.5$ and $\sigma = 0.1$, the EQ design with the misspecification on larger variance even works better than the EQ design with the correct specification and the ES design.  As we mentioned that the EQ method from the true prior does not count for the random effects between subjects, it is not surprising that the EQ design with the misspecified larger variance can worker better in some cases.     

In summary, the EQ design with the correctly specified prior distribution outperforms the ES design in most settings. Under mis-specification, the EQ design still outperforms the ES design, except for two scenarios: (1) when the center is at the symmetric center and the spread is big (i.e. $\mu=0.5$ and $\sigma = 0.1$); (2) when $\mu=0.75, \sigma=0.01$, and the mis-specified center is shifted to the left. For the former, it happens because the performance of the ES design improve a lot with a symmetric prior and also with priors with less information (larger variance). The latter happens because the EQ design schedules more visits on the left of the true center, which results fewer visits on the right side of the center and, therefore, causes trouble in the estimation. Example is shown in the top right corner of Figure \ref{fig:size10}, which illustrates all non-terminal time points (plus symbols) are scheduled on the left-hand side of the peak of the change-point distribution.

\section{Data Application}	

To make directly comparison on performance of the EQ and ES designs, we require data from studies, which have been carried out based on both designs simultaneously. Since, in practice, such studies have never been conducted, we propose an indirect comparison approach. First of all, we identify a study with three minimum requirements: (1) containing measurements of at least one cognitive function over time; (2) having a sufficient number of visits for each participant; and (3) containing multiple groups with potential within-group homogenous pattern between cognitive functions and age. Secondary, we focus on those homogenous groups. For each group, we fit a change-point model to measures of cognitive function and obtain the estimated standard deviation of the slope increasement, $\beta_{2}$. And we calculate the similarities between its actual design and the ES and EQ design separately. Thirdly, we investigate whether the similarities in designs are associated with the magnitude of the estimated variance of $\beta_{2}$. 

\begin{figure}[h!btp]
\begin{center}
\vspace{0.1in}
\includegraphics[scale=0.5]{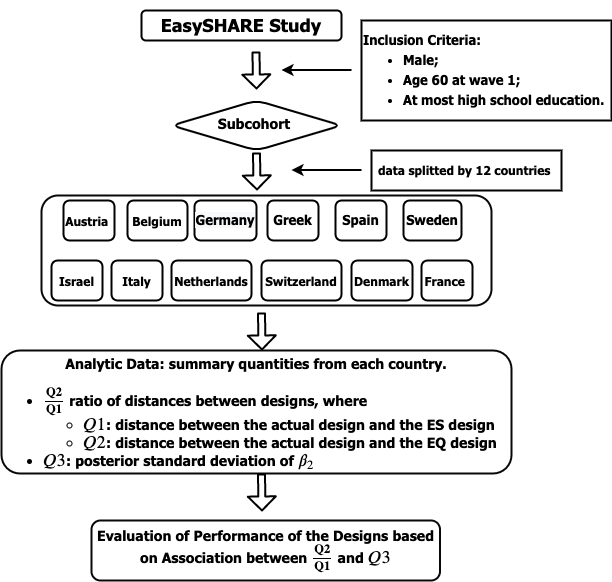}
\caption{Diagram of Procedure from Data Collection to Analysis.}  \label{fig:diag}
\end{center}
\end{figure}

The Survey of Health, Aging and Retirement in Europe (SHARE) study \cite{10.1093/ije/dyt088} is a multidisciplinary and cross-national panel database of micro data on health, socio-economic status and social and family networks of about 140,000 individuals aged 50 or older (around 380,000 interviews) and so far contains measurements on cognitive measurements at most $7$ waves for each participant \cite{SHAREw1, SHAREw2, SHAREw3, SHAREw4, SHAREw5, SHAREw6, SHAREw7}. Within the SHARE, the EasySHARE is a simplified HRS-adapted dataset, which is suitable to illustrate this indirect approach \cite{easySHARE2014, easySHARE2019}. Therefore, we start from the EasySHARE study and conduct the indirect comparison pipeline illustrated in Diagram~\ref{fig:diag}. 

As shown in the diagram, we only include male participants, who entered the study at wave 1, were aged $60$ years old, and also have at most high school education level. This decision is mainly due to the third requirement on groups with potential within-group homogenous pattern. A previous publication \cite{Dewey2005} reported that in SHARE, the main factors affecting cognitive function are age, gender, education, and country. Therefore, we give restrictions on age, gender, and education and would like to split the data by country to generate the desired groups. Under these inclusion criteria,  we have $1705$ samples in total. After that, we split the samples by $12$ countries (Israel and $11$ European countries). For each country, we calculate three key quantities, Q1, Q2, and Q3, where Q1 and Q2 are distance measurements between the actual design and the ES and EQ designs separately and Q3 is the estimated standard deviation of $\beta_{2}$ from the fitted change-point model. Here is more detail on how to calculate Q1. Within each country, for every participant, we treat that participant's times of visits as an empirical distribution of the actual design. Then we calculate the Kolmogorov–Smirnov (KS) distance \cite{doi:10.1080/01621459.1974.10480196, JSSv008i18} between this distribution and a uniform distribution (i.e. the ES design). Q1 is the average of those distance measures, which represents similarity between the actual design and the ES design for that country.  The larger Q1 corresponds to the greater dissimilarity. The calculation of Q2 is slightly complex. Since we assume that there is a homogenous pattern within each country, the change-points estimated based on each individual data are treated as samples from the true change-point distribution. We use those change-points as an empirical distribution of the true change-point distribution and then work out the PMF of the EQ design based on Algorithm~\ref{algm} given the weights as $(0, 0, 1)$. Then we calculate the KS distance between each participant's visits with the PMF of the EQ design. Q2 is the average of those quantities. Therefore, Q2 represents a similarity measurement between the actual design and the EQ design. At last, we conduct the Bayesian inference of the change-point model to each country and extract the posterior standard deviation of $\beta_{2}$, which corresponds to the scenario of giving $100$\% focus on the increment of the second slope.                     

\begin{figure}[h!btp]
\begin{center}
\vspace{0.1in}
\includegraphics[scale=0.7]{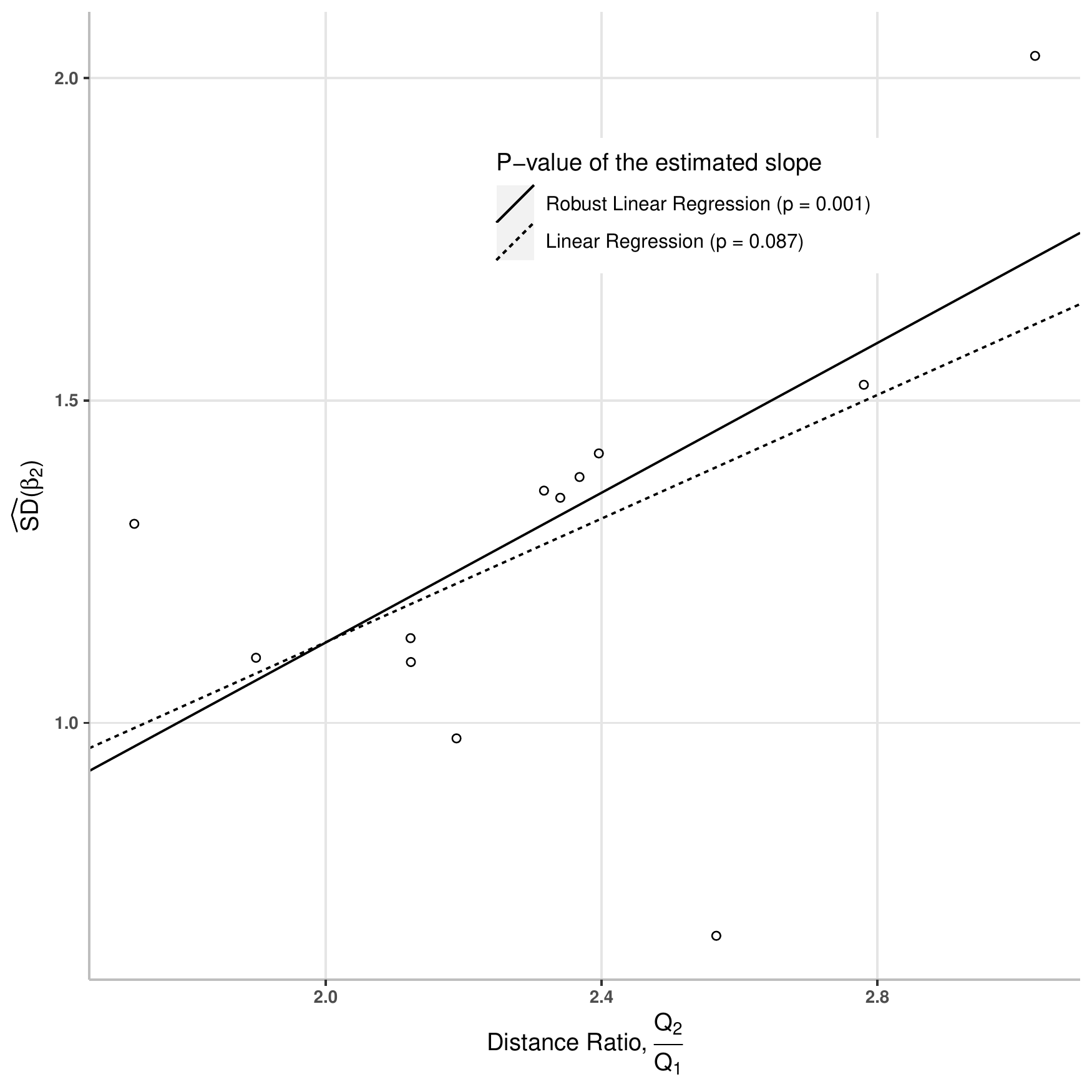}
\caption{Scatter plot of the EQ distance to ES distance ratio versus the estimated standard deviation of $\beta_{2}$. The dashed line is the fitted linear regression model. The solid line is the fitted robust linear regression model.}  \label{fig:trend}
\end{center}
\end{figure}
Figure~\ref{fig:trend} demonstrates a strong positive association between the EQ to ES distance ratio, $Q_{2}/Q_{1}$, and the estimated standard deviation of $\beta_{2}$. As shown in the figure, the p value for the estimated slope from the linear regression model is $0.087$, which is a little above the common significant threshold, $0.05$, due to the effect of some outlier.  And we also conduct the robust linear regression model to focus on the majority data. It results p-value$=0.001$ indicating a significant association. Therefore, we conclude that the more similar the actual design to the EQ design the better precision we have for the $\beta_{2}$ estimate.   
 
\section{Discussion and Conclusion}

The most important benefit of the proposed EQ design is to incorporate prior information of the change-point locations from historical data and incorporate it in the study design, which helps to better detect the pattern change. The traditional ES design, i.e. time scheduling on equal time intervals, can be treated as a special case of the EQ under a flat prior. Compared with the ES design, the EQ design provides more precise model parameter estimates given the same sample sizes. And the EQ design is also robust regards to certain degrees of mis-specification of the prior distribution. This design can be employed in other fields, where a change-point model is appropriate and researchers have prior info on the change-point location and would like to design new studies to get more precise estimate of the change-point.  

Another thing to point out is that the performance of the EQ design with mis-specified center to the right side reassure us the stability of the EQ design. Since there is always a time lag in real life, when we observe a change-point, it actually happened already. So most likely there is time lag bias in the change-point prior based on historical data. Our investigation show that the EQ\_Right designs are quite robust and maintain good performance under such bias.          

\section{Acknowledgements}
This research is supported by NSERC through the Discovery Grants program, through the Canada Research Chair program, and through the NSERC Postdoctoral Fellowships Program and by the University of Victoria through a UVic Internal Research Grant and the UVic Faculty of Science. This research is also supported by Integrative Analysis of Longitudinal Studies on Aging and Dementia (IALSA) research network. This research was enabled in part by support provided by WestGrid (www.westgrid.ca) and Compute Canada (www.computecanada.ca).

The SHARE data collection has been primarily funded by the European Commission through FP5 (QLK6-CT-2001-00360), FP6 (SHARE-I3: RII-CT-2006-062193, COMPARE: CIT5-CT-2005-028857, SHARELIFE: CIT4-CT-2006-028812) and FP7 (SHARE-PREP: N$^{o}$2-11909, SHARE-LEAP: N$^{o}$227822, SHARE M4: N$^{o}$261982). Additional funding from the German Ministry of Education and Research, the Max Planck Society for the Advancement of Science, the U.S. National Institute on Aging (U01\_AG09740-13S2, P01\_AG005842, P01\_AG08291, P30\_AG12815, R21\_AG025169, Y1-AG-4553-01, IAG\_BSR06-11, OGHA\_04-064,   
HHSN271201300071C) and from various national funding sources is gratefully acknowledged (see www.share-project.org).

At last, the authors would like to thank Professor Raj Srinivasan at the University of Saskatchewan and Professor Jeff Babb at the University of Winnipeg for constructive suggestions during preparation of this manuscript.

\bibliographystyle{plain}
\bibliography{ChangePointReference}

\end{document}